# Dynamic Fréchet Regression with Feature Selection for Distributional Data

Kiran Adhikari[a], Amrutha Dinesh[b], Mathew Kuttolamadom[b, c], Ying Lin[a, d*]

*[a] Industrial and Systems Engineering, University of Houston, Houston, USA; [b] Materials Science and Engineering, Texas A&M University, College Station, USA; [c] Engineering Technology and Industrial Distribution, Texas A&M University, College Station, USA; [d] Advanced Manufacturing Institute, University of Houston, Houston, USA*

*Corresponding Author: ylin58@uh.edu

# Dynamic Fréchet Regression with Feature Selection for Distributional Data

## Abstract

Many scientific and engineering applications generate responses that are not scalars or vectors, but statistical objects whose form evolves over an ordered index such as time, depth. Probability distributions are a prominent example, capturing variability and uncertainty that cannot be summarized by low-dimensional statistics. When such responses are observed sequentially, the resulting dynamic distributional trajectories pose significant challenges for regression, particularly in relating scalar predictors to both within-index variability and cross-index evolution. We propose Dynamic Fréchet Regression (DFR), a framework for modeling index-dependent trajectories of distribution-valued responses. DFR extends Global Fréchet Regression by introducing an index-aware weighting mechanism. At each index, predictions are defined as weighted Fréchet means in a metric space of distributions (e.g., Wasserstein space), preserving the intrinsic geometry of the response. The weights depend jointly on predictor similarity and index proximity, enabling index-specific prediction while borrowing strength across neighboring indices. To improve interpretability in high-dimensional settings, DFR incorporates a geometry-aware feature selection approach based on sparse metric learning, which identifies predictors driving distributional dynamics without relying on Euclidean coefficients. Simulation studies show improved predictive accuracy and feature recovery over existing methods. An application to additive manufacturing data demonstrates its ability to produce interpretable, index-specific distributional predictions.

## 1. Introduction

Many contemporary scientific and engineering problems produce data in which the response of interest is not a scalar or vector, but a statistical object whose form evolves over an index such as time, depth, or spatial layer (Cheng et al. 2021; Marron and Alonso 2014). Probability distributions are a prominent example: they capture intrinsic variability, uncertainty, and heterogeneity that cannot be adequately summarized by low-dimensional statistics. In many applications, these distribution-valued responses are observed sequentially, giving rise to dynamic distributional trajectories whose shapes and evolution are driven by a set of scalar predictors (Yang et al. 2019). Understanding how such predictors influence both the *within-index variability* and the *cross-index evolution* of distributions poses a fundamental challenge for regression methodology.

Additive manufacturing provides a representative motivating example. In layer-by-layer processes such as Directed Energy Deposition (DED), process parameters (e.g., laser power, scan speed, dwell time) are fixed at the build level, while system responses—such as melt pool area or temperature—are recorded repeatedly within each layer and across successive layers (Era et al. 2023; Qin et al. 2022). Due to inherent stochasticity in thermal dynamics, material flow, and measurement noise, melt pool behaviour exhibits substantial within-layer variability, making probability distributions a natural response representation (He et al. 2025). At the same time, thermal accumulation and material history induce systematic, smooth evolution of these distributions across layers, even under static process settings. Each experimental run therefore yields a layer-indexed sequence of distributions, whose trajectory depends on scalar process parameters (Zhang et al. 2021). Similar data structures arise in many other domains, including environmental monitoring, neuroscience, and reliability engineering (Cheng et al. 2021).

This setting highlights a broader methodological problem: how to regress dynamic, distribution-valued responses on scalar predictors in a way that is stable, interpretable, and respectful of the underlying geometry of the response space. Classical regression models, which assume Euclidean responses, are ill-suited for this task. Regressing distribution summaries such as means or variances discards essential information, while vectorizing distributions ignores their intrinsic structure. Functional data analysis (Morris 2015) provides tools for modeling responses that evolve smoothly over an index, but it typically operates in Euclidean spaces and therefore requires transforming distributions into vectors. Such linearizing transformations embed distributions into a vector space at the cost of distorting their natural geometry and are generally designed for single, univariate distributional responses rather than structured, index-dependent distributional trajectories (Han et al. 2020; Yang et al. 2019).

Recent advances in distributional regression extend classical regression to responses taking values in general metric spaces — commonly referred to as Fréchet regression (Petersen and Müller 2019) — by estimating conditional Fréchet means, including in the Wasserstein space of probability distributions (Dubey and Muller 2017). These methods are formulated for settings in which each predictor realization corresponds to a single response object and implicitly assume independence across observations. As a result, when applied to indexed distributional data, they treat responses at different indices as exchangeable and fail to exploit cross-index dependence or smooth evolution over indices. Moreover, prediction from these methods is driven solely by covariate similarity, without incorporating index-dependent weighting, which fails to infer dynamic evolution in distributional responses.

In addition, most existing distributional regression methods lack principled mechanisms for feature selection. In Euclidean regression, variable selection is achieved through shrinking coefficients, but for distribution-valued responses explicit coefficients are generally unavailable. To address this challenge, existing approaches regularize predictor-dependent weighting functions or employ marginal screening strategies and can achieve strong theoretical and empirical performance in static settings. Nevertheless, these strategies do not naturally extend to dynamic contexts as a result, identifying predictors that drive distributional dynamics remains an open challenge.

In summary, existing methods either respect distributional geometry but ignore dynamic structure, or model dynamics while relying on Euclidean assumptions that are incompatible with regression and feature selection for distribution-valued responses. A clear methodological gap therefore remains. Addressing this gap is essential for the principled modeling of dynamic distributional data in additive manufacturing and beyond.

To address these challenges, we propose Dynamic Fréchet Regression (DFR), a regression framework for modeling index-dependent evolution dynamics of distribution-valued responses from scalar predictors. DFR extends Global Fréchet Regression (GFR) by explicitly incorporating the evolution dynamics through an index-aware weighting mechanism that enables local, index-specific prediction while borrowing strength across neighboring indices. At each index, the predicted response is defined as a weighted Fréchet mean in the underlying metric space, ensuring that estimation and prediction respect the intrinsic geometry of probability distributions, such as the Wasserstein space. The weights are constructed to depend jointly on predictor similarity and index proximity, allowing DFR to capture smooth distributional evolution without imposing Euclidean linear structure on the response. This design yields stable predictions in noisy settings and produces coherent distributional trajectories that reflect systematic dynamic effects.

DFR further introduces a geometry-aware feature selection mechanism tailored to distribution-valued responses. Rather than relying on regression coefficients, predictor relevance is learned through a sparse metric that governs how scalar inputs influence the similarity weights defining the Fréchet mean. This approach enables identification of predictors that drive distributional dynamics while naturally excluding nuisance variables, even in high-dimensional and correlated predictor spaces. For univariate distributions, DFR admits an efficient implementation in quantile space: under the 2-Wasserstein metric, the estimation problem reduces to a weighted isotonic regression that guarantees valid distributional outputs and scales efficiently to practical data sizes. Through simulation studies and an additive manufacturing case study, we demonstrate that DFR achieves superior predictive accuracy, improved stability, and enhanced interpretability compared with existing distributional and functional regression methods.

The remainder of the paper is organized as follows. Section 2 reviews related work on distributional regression, scalar-to-function modeling, and feature selection. Section 3 introduces the proposed Dynamic Fréchet Regression (DFR) framework, including its mathematical formulation, index-aware weighting mechanism, and geometry-aware feature selection via sparse metric learning. Section 4 presents simulation studies comparing DFR with Global Fréchet Regression (GFR) and Function-on-Scalar Regression (FoSR), demonstrating its ability to model dynamic distributional responses and recover active predictors in the presence of nuisance variables. Section 5 applies DFR to experimental melt pool data from a Directed Energy Deposition (DED) process, illustrating its capacity to produce interpretable, layer-specific distributional effects of process parameters. Section 6 concludes with a summary of contributions and directions for future research.

## 2. Literature Review

To motivate Dynamic Fréchet Regression (DFR) and its geometry-aware feature selection mechanism, we review related work in three areas: (i) scalar-to-function regression for index-dependent responses, (ii) statistical methods for distribution-valued data, and (iii) feature selection algorithms, particularly in non-Euclidean and structured-response settings.

### *2.1 Scalar-to-Function Regression*

Scalar-to-function regression, also known as Function-on-Scalar Regression (FoSR), provides a foundational framework for modeling responses that vary smoothly over an index such as time, space, or depth while depending on scalar predictors. A comprehensive overview of this area is given by Morris (2015) who highlights functional data analysis (FDA) as a paradigm that treats each response curve as a single structured object and emphasizes the dual roles of replication across functions and regularization within functions through basis representations such as splines, wavelets, and functional principal components. Building on this foundation, Bauer et al. (2018) introduced a flexible semiparametric function-on-scalar regression framework based on generalized additive models and penalized splines, providing practical guidance for modeling complex functional responses, including time series and spatially indexed functions. Several works have extended scalar-to-function regression to distributional responses by embedding distributions into Euclidean function spaces. Yang et al. (2019) proposed quantile function-on-scalar regression, modeling subject-specific distributions via their quantile functions and employing custom basis expansions and Bayesian shrinkage to identify covariate effects on distributional features. Similarly, Han et al. (2020) introduced additive functional regression for density-valued responses by transforming densities into square-integrable functions through log-quantile and log hazard transformation. While these approaches effectively leverage functional regression machinery and allow flexible modeling of distributional responses, they rely on transformations that linearize distributions and thus do not preserve the intrinsic geometry of probability spaces. Moreover, they are not explicitly designed to handle multivariate distributions and index-dependent sequences of distributions with strong cross-index dependence. These limitations motivate us to develop a novel regression framework that retains the strengths of scalar-to-function modeling —namely, smooth evolution across the index and modeling of cross-index dependence within the response — while operating directly on distribution-valued responses in their native metric spaces.

### *2.2 Distributional Data Analysis*

A variety of methodological frameworks have been proposed for regression with non-Euclidean responses. An early and general strategy is distance-based regression, as proposed by Faraway (2014) which represents complex objects—such as shapes, curves, and correlation matrices—via distance matrices. While broadly applicable, such approaches rely on complicated scoring and back scoring steps. A more geometry-preserving framework is provided by Fréchet regression, introduced by Petersen and Müller (2019), which extends classical regression to metric-space–valued responses by defining prediction as conditional Fréchet mean. This framework accommodates responses such as probability distributions and correlation matrices, and offers theoretical guarantees under suitable regularity conditions. Subsequent work has expanded Global Fréchet Regression to address practical challenges, including model uncertainty and flexibility. Yan, Zhang, and Zhao (2025) proposed a frequentist model averaging approach for regression of density responses, selecting model weights via cross-validation under the Wasserstein distance to improve predictive performance and robustness to misspecification. More recently, Iao, Zhou, and Müller (2025) introduced Deep Fréchet Regression, combining manifold learning and deep neural networks to address high-dimensional predictors and nonlinear structure while mapping predictions back to the original metric space via local Fréchet regression. In dynamic settings, several works have been developed under the Wasserstein metric. Zhang, Kokoszka, and Petersen (2020) proposed Wasserstein autoregressive (WAR) models for density time series, exploiting the tangent space structure induced by the Wasserstein geometry to define autoregressive dynamics directly on the space of probability densities. Their order-p Wasserstein AR model enables flexible, nonparametric forecasting of serially dependent densities without imposing parametric assumptions on the density family. Similarly, Cheng et al. (2022) proposed nonparametric and regularized dynamical Wasserstein barycenter's for sequential observations, modeling time-evolving distributions as a Wasserstein barycenter of a finite set of pure-state distributions with time-varying barycentric weights. While these approaches effectively model temporal dependence among distributions, they do not explicitly incorporate scalar covariates that may explain or drive the evolution of the distributional trajectory nor do they provide mechanisms for interpretable feature selection. In summary, existing methods do not adequately address regression settings as pursued in this paper.

### *2.3 Feature Selection*

Zhang, Xue, and Li (2024) proposed a sufficient dimension reduction framework for Fréchet regression that maps metric-space–valued responses, such as probability distributions, into scalar similarity measures using kernel functions based on distance metrics and anchor distributions, and then applied classical sufficient dimension reduction methods to recover low-dimensional predictor subspaces that preserve regression information. However, when the response consists of index-dependent sequences of distributions corresponding to each predictor vector, this scalarization is not straightforward, as distance between two entire distributional trajectories is not well-defined. Tian, Kang, and Zhong (2025) introduced Fréchet-SIS, a sure independence screening procedure based on marginal Global Fréchet Regression, providing theoretical guarantees for retaining all relevant predictors with high probability. Nevertheless, as a marginal and index-independent method, Fréchet-SIS ignores joint predictor effects. Tucker, Wu, and Müller (2023) introduced the variable selection framework for Global Fréchet Regression by leveraging how predictors influence the similarity weights defining the Fréchet mean. Their method, FRiSO, incorporates predictor-specific ridge regularization within a Mahalanobis-type similarity metric that determines predictor-dependent weights. By increasing the penalty associated with uninformative predictors, their influence on the similarity weights is effectively diminished. Through a constrained optimization over the sum of the inverses of the ridge regularization parameters, FRiSO shrinks cenrtain rows and columns of weight matrix to zero, thereby yielding sparse predictor sets and establishing selection consistency with strong finite-sample performance. However, a key limitation of FRiSO is that the similarity weights depend on the inverse of the sample covariance matrix, which is fixed and not learned from the regression objective. Consequently, when predictors are weakly correlated or uncorrelated, the covariance structure provides little information about their joint relevance. Consequently, it will limit effectiveness of FRiSO in predictor selection when the response depends on interaction effects apart from presence or absence of marginal effects. In light of these limitations**,** feature selection can be naturally incorporated into Fréchet regression by embedding sparse metric learning directly into the similarity-based weighting mechanism. Huang and Sun (2013) proposed a sparse metric learning method for feature selection in kernel regression, in which a Mahalanobis distance with group-type regularization simultaneously induces sparsity, performs dimension reduction, and governs similarity-based weighting. However, existing sparse metric learning methods are primarily developed for Euclidean responses and do not account for index-aware structure or geometry-preserving prediction in metric spaces. In this work, we extend this idea to Dynamic Fréchet Regression by integrating sparse metric learning into an index-aware weighting mechanism, enabling principled feature selection for dynamically evolving distribution-valued responses.

## 3. Methodology

### *3.1 Problem Formulation*

Assume a training dataset $D = \{\boldsymbol{x_i}, \boldsymbol{Y}_i\}_{i=1}^{N}$, where each observation includes a vector of scalar predictor $\boldsymbol{x_i} \in \mathbb{R}^p$ and a sequence of metric-valued responses $\boldsymbol{Y}_i = (Y_{i1}, \dots \dots, Y_{iL}) \in \mathcal{M}^L$, where each $Y_{il}$ is a probability distribution belonging to a metric space $(\mathcal{M}, d)$ and $d(\cdot\,,\ \cdot)$ measures discrepancy between distributions (e.g., the 2-Wasserstein distance). The index $l$ represents an ordered dimension (e.g., time, depth, or layer), and distributions are assumed to evolve smoothly across $l$. The objective is to learn a regression map, $f(\boldsymbol{x})\colon \mathbb{R}^p \to \mathcal{M}^L$, that predicts the sequence of metric-valued responses $\{Y_{0l}\}_{l=1}^{L}$ for a new predictor value $\boldsymbol{x}_0$, while respecting the geometry of $\mathcal{M}$ and borrowing strength across indices.

### *3.2 Global Fréchet Regression*

Existing distributional regression models are motivated by the observation that prediction at a new predictor value can be formulated as the solution to a weighted distance-minimization problem over the response space. Specifically, given training responses $\{Y_i\}_{i=1}^{n}$, a generic regression predictor can be written as

$$f(\boldsymbol{x}_0) \;=\; \arg\min_{\omega \in \Omega} \sum_{i=1}^{n} w_i(\boldsymbol{x}_0)\, d^2(Y_i, \omega) \tag{1}$$

where $d(\cdot,\cdot)$ is a distance on the response space $\Omega$ and $w_i$ is a nonnegative weight that quantifies the relevance of the $i$th training sample to the target predictor value, $\boldsymbol{x}_0$.

In Euclidean settings, linear regression and kernel regression can both be expressed in this form, with weights ($w_i$) determined by linear projection or kernel similarity, respectively, and with squared Euclidean distance used to measure discrepancies between responses, i.e. $d^2(Y_i, \omega)$. These models rely fundamentally on the linear structure of the response space. To accommodate metric-valued responses, Global Fréchet Regression generalizes this principle by replacing the Euclidean distance with a metric appropriate for the response space, such as the Wasserstein distance for probability distributions (Petersen and Müller 2019). The resulting estimator is a weighted Fréchet mean of the training outcomes, providing a geometry-preserving extension of classical regression to distribution-valued data.

Although Global Fréchet Regression (GFR) preserves the geometry of metric-valued responses, it is inherently static. The weights in GFR depend only on predictor similarity and are constant across indices. When responses are observed as indexed sequences $\{Y_{il}\}_{l=1}^{L}$, the model formulation in Eq. (1) treats all responses as exchangeable and ignores the index-dependent varying effects of predictors. As a result, GFR cannot capture systematic evolution across indices or borrow strength locally from neighboring index locations, leading to unstable predictions in noisy settings.

### *3.3 Dynamic Fréchet Regression*

The proposed Dynamic Fréchet Regression (DFR) model extends the formulation in Eq. (1) to predict index-specific metric-valued responses by introducing an index-aware weighting function. Specifically, for a new predictor value $\boldsymbol{x}_0$ and target index $j$, the DFR estimator is defined as:

$$f(\boldsymbol{x}_0, j) = \underset{\omega \in \Omega}{\arg\min} \sum_{i=1}^{N} \sum_{l=1}^{L} w_{il}(\boldsymbol{x}_0, j) d^2(Y_{il}, \omega) \tag{2}$$

where $f(x_0, j)$ denotes the predicted distribution at index $j$, and $w_{il}(\boldsymbol{x}_0, j)$ is a nonnegative weight that depends jointly on predictor similarity and index proximity. Similar to Global Fréchet Regression, DFR generalizes the Euclidean average to non-Euclidean spaces by selecting the distribution $\omega \in \Omega$ that is "closest on average" to the training responses $\{Y_{il}\}$, where closeness is measured by the squared Wasserstein distance $d^2(\cdot,\cdot)$. Unlike GFR, however, DFR employs index-aware weights to distinguish (i) the relevance of training observations observed at different indices and (ii) the index-specific nature of the prediction. As a result, the DFR estimator can be viewed as a geometry-aware average that emphasizes training distributions most relevant to both the target predictor value $x_0$ and the target index $j$.

### *3.4 Index-aware Weighting Function*

A central challenge in DFR is the construction of an effective index-aware weighting function that balances local index specificity with stable borrowing of information across neighboring indices, while remaining responsive to predictor similarity. Motivated by the fact that the weighting function of the GFR can be obtained from the least square estimator of linear regression, we analyze the least-squares estimator of function-on-scalar regression and show how its smoothing matrix induces index-dependent weights. Leveraging this insight, we develop a novel index-aware weighting function for DFR that extends Global Fréchet Regression to dynamic, index-indexed settings in a theoretically grounded and interpretable manner.

To motivate the construction of the index-aware weighting function in DFR, we begin by noting a useful property of function-on-scalar regression in a Euclidean setting: its prediction at a target index can be expressed as the minimizer of a weighted sum of squared distances. This representation provides a natural bridge between functional regression and Fréchet regression.

Specifically, consider the special case in which the response space is Euclidean, $\Omega^L = \mathbb{R}^L$. Let $z_{il}$ denote a functional response for subjects $i = 1, \dots, N$ observed at ordered index $l = 1, \dots, L$.

*Property 1: Under roughness-penalized function-on-scalar regression model (Reiss et al. 2010), the predicted response at a new predictor value $\boldsymbol{x}_0$ and target index $j$ can be expressed as the minimizer of a weighted sum of squared Euclidean distances:*

$$\hat{z}(\boldsymbol{x}_0, j) = arg \min_{u \in \mathbb{R}} \sum_{i=1}^{i=N} \sum_{l=1}^{L} w_{il}(\boldsymbol{x}_0, j)(z_{il} - u)^2 \quad (3)$$

*where the weights $w_{il}(\boldsymbol{x}_0, j)$ depends on both predictor similarity and index proximity. When B-spline functions are used to represent the functional response and a roughness penalty is imposed, the index-aware weights admit an explicit matrix form:*

$$w_{il}(\boldsymbol{x}_0, j) = \left[\left(\boldsymbol{\Theta}(\boldsymbol{\Theta}^T\boldsymbol{\Theta})^{-1}\boldsymbol{\Theta}^T + (\boldsymbol{\Theta} \otimes (\boldsymbol{x}_0 - \overline{\boldsymbol{x}})^T)\left[\boldsymbol{\Theta}^T\boldsymbol{\Theta} \otimes \hat{\Sigma} + \lambda * \boldsymbol{R} \otimes diag(p)\right]^{-1}(\boldsymbol{\Theta}^T \otimes (\boldsymbol{x}_i - \overline{\boldsymbol{x}}))\right)\right]_{jl} \quad (4)$$

This property shows that prediction at new predictor value and specific index, $\hat{z}(\boldsymbol{x}_0, j)$, is obtained as a weighted aggregation of all observed responses, with the weights determining the relative contribution of subject $i$ and index $l$ to the target $(\boldsymbol{x}_0, j)$. In the weighting function, $\boldsymbol{x}_i \in \mathbb{R}^p$ denotes the predictor vector for the $ith$ training subject and $\bar{x} = \frac{1}{N}\sum_{i=1}^{N} \boldsymbol{x}_i$ is the sample mean of the predictors; the centered quantities $\boldsymbol{x}_0 - \overline{\boldsymbol{x}}$ and $\boldsymbol{x}_i - \overline{\boldsymbol{x}}$ ensure that the covariate-driven contribution to the weights is expressed through deviations from the global predictor mean. The matrix $\widehat{\boldsymbol{\Sigma}} \in \mathbb{R}^{p \times p}$ is the covariance matrix of $\{\boldsymbol{x}_i\}_{i=1}^{N}$, which scales predictor differences in the covariate-dependent term of the smoother. Moreover, $\boldsymbol{\Theta} \in \mathbb{R}^{L \times K}$ is the B-spline basis matrix evaluated on the common grid $\{t_l\}_{l=1}^{L}$, and $\boldsymbol{R} \in \mathbb{R}^{K \times K}$ is the corresponding spline roughness penalty matrix constructed from integrated squared second derivatives of the basis functions. The roughness parameter $\lambda \geq 0$ controls the extent of cross-index smoothing through the penalized inverse $[\boldsymbol{\Theta}^T \boldsymbol{\Theta} \otimes \widehat{\sum} + \lambda * \boldsymbol{R} \otimes diag(p)]^{-1}$, where $diag(p)$ denotes the $p \times p$ diagonal matrix.

Property 1 provides the conceptual foundation for Dynamic Fréchet Regression. It reveals that function-on-scalar regression prediction can be viewed as minimizing a weighted average of squared distances, with index-aware weights that balance local fidelity and smoothness. DFR extends this representation to distribution-valued responses by replacing the squared Euclidean distance in (3) with a squared metric distance $d^2(\cdot,\cdot)$ on a general metric space $\Omega$, and by replacing the scalar decision variable $u \in \mathbb{R}$ with a candidate distribution $\omega \in \Omega$. The resulting estimator is therefore a weighted Fréchet mean with index-aware weights, yielding the DFR predictor in Eq. (2).

The roughness parameter $\lambda$ plays a critical role in DFR: small values favor locally adaptive, index-specific predictions, while larger values promote smoother weighting patterns and stronger borrowing of information across indices. From the weight construction in Eq. (4), increasing $\lambda$ shrinks the influence of the covariate-driven correction term: $[\boldsymbol{\Theta}^T \boldsymbol{\Theta} \otimes \widehat{\sum} + \lambda * \boldsymbol{R} \otimes diag(p)]^{-1}$, thereby reducing sensitivity to predictor differences across indices and yielding smoother distributional trajectories. In the limit as $\lambda \to 0$, DFR behaves as a locally adaptive, index-specific Fréchet regression, whereas as $\lambda \to \infty$, the weights are dominated by the spline smoother and become largely independent of predictors.

### *3.5 Feature Selection in Dynamic Fréchet Regression*

Identifying predictors that drive distributional dynamics is nontrivial in Dynamic Fréchet Regression, because the model does not involve explicit regression coefficients. As a result, standard coefficient-based regularization and selection methods are not directly applicable.

Instead, prediction in DFR is entirely determined by the similarity weights $w_{il}(\boldsymbol{x}_0, j)$. This observation motivates a different notion of predictor relevance: a predictor is important if it meaningfully influences how the similarity weights vary with the input $\boldsymbol{x_0}$, and therefore affects which training distributions are emphasized in prediction. To formalize this idea, we consider a simplified setting with $\lambda = 0$, which removes the roughness penalization while preserving the spline-based smoothing across indices through the basis matrix $\boldsymbol{\Theta}$. Under this setting, the index-aware weights admit an explicit representation that provides a principled foundation for learning a sparse similarity metric and performing feature selection. Specifically, the weights in Eq. (4) reduce to:

$$w_{il}(\boldsymbol{x}_0, j) = \left[[\boldsymbol{\Theta}(\boldsymbol{\Theta}^T\boldsymbol{\Theta})^{-1}\boldsymbol{\Theta}^T]\left[1 + (\boldsymbol{x}_0 - \overline{\boldsymbol{x}})^{\boldsymbol{T}}\, \widehat{\Sigma}^{-\mathbf{1}}\, (\boldsymbol{x_i} - \overline{\boldsymbol{x}})\right]\right]_{jl} \tag{5}$$

where $\widehat{\Sigma}^{-\mathbf{1}}$ is the empirical precision matrix of the predictors. Equation (5) shows that predictors influence the weights only through the bilinear similarity score $1 + (\boldsymbol{x}_0 - \overline{\boldsymbol{x}})^{\boldsymbol{T}}\, \widehat{\Sigma}^{-\mathbf{1}}\, (\boldsymbol{x_i} - \overline{\boldsymbol{x}})$, which determins the covariate-based similarity between $\boldsymbol{x_i}$ and $\boldsymbol{x}_0$. This representation suggests a natural strategy for feature selection: rather than fixing the precision matrix $\widehat{\Sigma}^{-\mathbf{1}}$, we replace it with a learnable symmetric and positive semidefinite metric matrix $\boldsymbol{M} \succcurlyeq 0$. The matrix $\boldsymbol{M}$ determines how predictor differences are weighted when measuring similarity, thereby controlling which predictors influence the DFR weights. Specifically, we define

$$w_{il}(\boldsymbol{x}_0, j) = \left[[\boldsymbol{\Theta}(\boldsymbol{\Theta}^T\boldsymbol{\Theta})^{-1}\boldsymbol{\Theta}^T][1 + (\boldsymbol{x}_0 - \overline{\boldsymbol{x}})^{\boldsymbol{T}}\boldsymbol{M}\, (\boldsymbol{x_i} - \overline{\boldsymbol{x}})]\right]_{jl} \tag{6}$$

Under this formulation, the structure of $\boldsymbol{M}$ directly governs predictor relevance in the following property.

*Property 2: If the $k$th row (and column) of $\boldsymbol{M}$ is zero, then the $k$th predictor does not contribute to the similarity score and has no effect on the DFR weights defined in Eq. (6).*

Substituting the metric-dependent weights in Eq. (6) into the DFR estimator yields the metric-weighted predictor

$$g(\boldsymbol{x}_0, j|\boldsymbol{M}) = \underset{\omega \in \Omega}{\arg\min} \sum_{i=1}^{N} \sum_{l=1}^{L} w_{il}(\boldsymbol{x}_0, j; \boldsymbol{M}) d^2(Y_{il}, \omega) \tag{7}$$

To learn the similarity metric $\boldsymbol{M}$ , we evaluate how well the DFR model with metric $\boldsymbol{M}$ reconstructs the responses in training data. For each training pair $(\boldsymbol{x_i}, Y_{il})$, we compute the predicted distribution $g(\boldsymbol{x_i}, l|\boldsymbol{M})$ and measure its discrepancy from the observed response using the squared metric distance $d^2(\cdot,\cdot)$. Aggregating these discrepancies across all subjects and indices yields the empirical Fréchet reconstruction loss:

$$L(\boldsymbol{M}) = \frac{1}{N} \sum_{i=1}^{N} \sum_{l=1}^{L} d^2\big(Y_{il}, g(\boldsymbol{x_i}, l|\boldsymbol{M})\big),$$

To encourage predictor-level sparsity, we augment this loss with a group-lasso penalty and solve the following sparse metric learning problem:

$$\widehat{\boldsymbol{M}} = arg \min_{\boldsymbol{M} \succcurlyeq 0\,;\, \boldsymbol{M} = \boldsymbol{M}^T} \left\{ L(\boldsymbol{M}) + \mu \sum_{k=1}^{p} \left\| \boldsymbol{M}_{j,:} \right\|_2 \right\} \tag{8}$$

where $\boldsymbol{\mu} > \boldsymbol{0}$ controls the strength of sparsity. The group-lasso penalty $\sum_{k=1}^{\boldsymbol{p}} \left\| \boldsymbol{M}_{\boldsymbol{j},:} \right\|_{\boldsymbol{2}}$ shrinks entire rows (and, by symmetry, columns) of $\boldsymbol{M}$ toward zero, thereby removing predictors that do not contribute to the similarity structure underlying DFR. We solve the sparse metric learning problem in Eq. (8) using a gradient-based optimization scheme, parameterizing $\boldsymbol{M} = \boldsymbol{B}^T \boldsymbol{B}$ to enforce positive semidefiniteness. In the special case of univariate distributional responses as illustrated in Section 3.6, the squared Wasserstein loss admits a closed-form expression in quantile space, allowing the objective to be evaluated as a weighted squared $L^2$ loss between transformed quantile functions, which enables efficient gradient computation and stable optimization.

We declare predictor $k$ to be active if its row norm exceeds a small threshold $\tau > 0$ and the active set of predictors can be represented as:

$$\hat{S} = \left\{k: \left\|\widehat{\boldsymbol{M}}_{k,:}\right\|_2 > \boldsymbol{\tau}\right\} \tag{9}$$

Our estimation procedure proceeds in two stages. In the first stage, we perform feature selection by learning a sparse predictor metric from Eq. (8) while fixing $\lambda = 0$, so that selection is driven purely by covariate similarity in the weights. After identifying the active set $\hat{S}$, we refit DFR using only the selected predictors and then tune the roughness parameter $\lambda$ via cross-validation to control the degree of information sharing across indices. This separation yields a stable and interpretable model in which predictor relevance and index-wise smoothing are learned independently.

### *3.6 Application to Univariate Distribution*

When each response $Y_{il}$ is a univariate probability distribution, DFR admits an efficient implementation by representing distributions via their quantile functions. Let $\mathbb{Q} = \{q: (0,1) \rightarrow \mathbb{R} \text{ nondecreasing}\}$ denote the space of valid quantile functions, and let $Q: \Omega \rightarrow \mathbb{Q}$ be the mapping from a distribution to its quantile function. For $Y_{il} \in \Omega$, we write its quantile function as $q_{il} = Q(Y_{il})$ and for a candidate distribution $\omega \in \Omega$, we write $q = Q(\omega)$. The inverse map $Q^{-1}$ recovers a distribution from a valid quantile function. For univariate distributions, the 2-Wasserstein distance admits a closed-form expression in terms of quantile functions. Specifically, for $Y, \omega \in \Omega$ with quantile functions $q_Y$ and $q_\omega$

$$d^2(Y, \omega) = \int_0^1 \left(q_Y(u) - q_\omega(u)\right)^2 du \tag{10}$$

Thus, the prediction of a new predictor $\boldsymbol{x}_0$ and target index $j$ from DFR is simplified to a weighted Fréchet mean problem in quantile space,

$$f(\boldsymbol{x}_0, j) = Q^{-1}\left(\underset{q \in \mathbb{Q}}{\operatorname{argmin}} \sum_{i=1}^{N} \sum_{l=1}^{L} w_{il}(\boldsymbol{x}_0, j)\, d^2(q_{il}, q)\right) \tag{11}$$

where $w_{il}(\boldsymbol{x}_0, j)$ is the index-aware weight defined in Section 3.4, and depends jointly on predictor similarity and index proximity. In practice, we evaluate quantile functions on a fixed grid $0 < u_1 < \cdots < u_D < 1$ and approximate the distance metric $d^2(q_{il}, q)$ by a discrete sum $\sum_{d=1}^{D}\{q_{il}(u_d) - q(u_d)\}^2$, leading to the monotone weighted least-squares problem.

$$q(\boldsymbol{x}_0, j) = \underset{q(u_1) \le \cdots \le q(u_D)}{\operatorname{argmin}} \sum_{i=1}^{N} \sum_{l=1}^{L} w_{il}(\boldsymbol{x}_0, j) \sum_{d=1}^{D}\{q_{il}(u_d) - q(u_d)\}^2 \tag{12}$$

Define the total weight and the index-aware weighted average quantiles as:

$$w_{tot}(\boldsymbol{x}_0, j) = \sum_{i=1}^{N} \sum_{l=1}^{L} w_{il}(\boldsymbol{x}_0, j);\; z(\boldsymbol{x}_0, j, u_d) = \frac{1}{w_{tot}} \sum_{i=1}^{N} \sum_{l=1}^{L} w_{il}(\boldsymbol{x}_0, j) q_{il}(u_d) \tag{13}$$

Then Eq. (12) is equivalent to a weighted isotonic regression of $\{z(\boldsymbol{x}_0, j, u_d)\}_{d=1}^{D}$:

$$q(\boldsymbol{x}_0, j) = \underset{q(u_1) \le \cdots \le q(u_D)}{\operatorname{argmin}} \sum_{d=1}^{D} w_{tot}(\boldsymbol{x}_0, j)\{z(\boldsymbol{x}_0, j, u_d) - q(u_d)\}^2 \tag{14}$$

This isotonic regression can be solved efficiently using the Pool Adjacent Violators Algorithm (PAVA) (Busing 2022). The predicted distribution at index $j$ is then

$$f(\boldsymbol{x}_0, j) = Q^{-1}\big(q(\boldsymbol{x}_0, j)\big) \tag{15}$$

This quantile-based representation also enables an efficient implementation of the feature selection procedure for the univariate case. When incorporating the learnable similarity metric $\boldsymbol{M}$, the prediction at $(\boldsymbol{x}_0, j)$ becomes

$$q(\boldsymbol{x}_0, j|\boldsymbol{M}) = \underset{q(u_1)\leq\cdots\leq q(u_D)}{\operatorname{argmin}} \sum_{d=1}^{D} w_{tot}(\boldsymbol{x}_0, j; \boldsymbol{M})\{z(\boldsymbol{x}_0, j, u_d; \boldsymbol{M}) - q(u_d)\}^2$$

where the weights and weighted averages now depend on the learned metric $\boldsymbol{M}$ through the similarity function defined in Section 3.5. Using the closed-form expression of the 2-Wasserstein distance in quantile space, the empirical reconstruction loss for metric learning can be evaluated directly as a squared $L^2$ distance between quantile functions on the discretized grid. The sparse similarity metric is then estimated by solving

$$\widehat{\boldsymbol{M}} = arg \min_{\boldsymbol{M}\succcurlyeq 0\,;\, \boldsymbol{M} = \boldsymbol{M}^T} \left\{\frac{1}{N}\sum_{i=1}^{N}\sum_{l=1}^{L} d^2\big(q_{il}, q(\boldsymbol{x_i}, l|\boldsymbol{M})\big) + \sum_{k=1}^{p}\left\|\boldsymbol{M}_{k,:}\right\|_2\right\} \tag{16}$$

where the reconstruction loss in Eq. (16) is denoted by:

$$L(\boldsymbol{M}) = \frac{1}{N}\sum_{i=1}^{N}\sum_{l=1}^{L} d^2\big(q_{il}, q(\boldsymbol{x_i}, l|\boldsymbol{M})\big),$$

$L(\boldsymbol{M})$ can be further evaluated efficiently in Eq. (16) after log hazard transformation of quantile function.

$$\Psi(q(u_d)) = log\left(\frac{f(q(u_d))}{1-F(q(u_d))}\right),$$

where $f$ and $F$ denote the probability density and cumulative distribution functions, respectively.

This transformation embeds quantile functions into an unconstrained vector space. We denote the transformed quantiles by: $q_{\Psi}(u_d) = \Psi\big(q(u_d)\big)$ and $q_{il,\Psi}(u_d) = \Psi(q_{il}(u_d))$. Under this representation, the metric-weighted average $z_{\Psi}(\boldsymbol{x_i}, j, u_d; \boldsymbol{M})$ at quantile level $u_d$ becomes

$$z_{\Psi}(\boldsymbol{x_i}, j, u_d; \boldsymbol{M}) = \frac{\sum_{i=1}^{N}\sum_{l=1}^{L} w_{il}(\boldsymbol{x_i}, j; \boldsymbol{M}) q_{il,\Psi}(u_d)}{\sum_{i=1}^{N}\sum_{l=1}^{L} w_{il}(\boldsymbol{x_i}, j; \boldsymbol{M})}$$

Using the transformed representation sparse similarity metric is estimated by solving:

$$\widehat{\boldsymbol{M}} = arg \min_{\boldsymbol{M} \succcurlyeq 0\,;\, \boldsymbol{M} = \boldsymbol{M}^T} \left\{ \frac{1}{N} \sum_{i=1}^{N}\sum_{j=1}^{L}\sum_{d=1}^{D} \left\| z_{\Psi}(\boldsymbol{x_i}, j, u_d; \boldsymbol{M}) - q_{ij,\Psi}(u_d) \right\|_2^2 + \sum_{k=1}^{p} \left\| \boldsymbol{M}_{k,:} \right\|_2 \right\} \tag{17}$$

The reconstruction loss in Eq. (17) reduces to a squared Euclidean loss. We solve the sparse metric learning problem using a gradient-based optimization scheme (Huang and Sun 2013). To enforce the positive semidefinite constraint, we parameterize: $\boldsymbol{M} = B^T B$, where $B \in \mathbb{R}^{p \times p}$ is an unconstrained matrix. The overall algorithm is summarized in Table 1.

Table 1. Sparse Metric Learning Algorithm for Dynamic Fréchet Regression

**Input**: Training data $\{\boldsymbol{x}_i, Y_{il}\}$; $i = 1, \ldots, N$; $l = 1, \ldots, L$; sparsity parameter $\mu$; learning rate $\eta$; convergence tolerance $\epsilon$; selection threshold $\tau$

**Output**: Estimated sparse metric $\widehat{\boldsymbol{M}}$; selected predictor set $\hat{S}$

**Initialization:**

- Initialize $B \in \mathbb{R}^{p \times p}$ randomly
- Set iteration counter $t \leftarrow 0$
- Compute initial metric: $\boldsymbol{M}^{(\boldsymbol{0})} \leftarrow B^T B$
- Compute Initial loss: $\boldsymbol{L}^{(\boldsymbol{0})} \leftarrow L(\boldsymbol{M}^{(\boldsymbol{0})})$

**Repeat until convergence:**

- Update metric: $\boldsymbol{M}^{(\boldsymbol{t})} \leftarrow B^T B$
- Compute transformed weighted averages $z_{\Psi}(\tau_d)$.
- Evaluate reconstruction loss using Equation (17)
- Add sparsity penalty: $\boldsymbol{L}^{(\boldsymbol{t})} \leftarrow \frac{1}{N} L\big(\boldsymbol{M}^{(\boldsymbol{t})}\big) + \mu \sum_{k=1}^{p} \left\| \boldsymbol{M}^{(\boldsymbol{t})}{}_{k,:} \right\|_2$
- Update parameter matrix: $B \leftarrow B - \eta \nabla_B \boldsymbol{L}^{(\boldsymbol{t})}$
- Check convergence: if $\frac{|\boldsymbol{L}^{(\boldsymbol{t})} - \boldsymbol{L}^{(\boldsymbol{t-1})}|}{\boldsymbol{L}^{(\boldsymbol{t-1})}} < \epsilon$, stop
- Set iteration counter: $t \leftarrow t + 1$

**Feature Selection:**

- Compute final: $\widehat{\boldsymbol{M}} = B^T B$
- Define $\hat{S} = \left\{ \text{k}: \left\| \widehat{\boldsymbol{M}}_{k,:} \right\|_2 > \tau \right\}$

**Refitting:**

- The final DFR model is then refit using only predictors in $\hat{S}$

## 4. Simulation Studies

We conduct simulation studies to evaluate the proposed Dynamic Fréchet Regression (DFR) framework for predicting sequences of index-specific probability distributions from static predictors. The simulations are designed to capture the two key structures targeted by DFR: (i) covariate-driven variation in distributions, and (ii) smooth evolution across indices, whereby distributions at neighboring indices are similar. To assess the effectiveness of the proposed geometry-aware feature selection procedure, we consider high-dimensional settings that include noisy predictors, allowing us to evaluate DFR's ability to accurately recover the true active predictors.

### *4.1 Simulation Models*

Each simulated dataset consists of 50 ($N = 50$) independent samples, where sample $i$ is characterized by a vector of scalar predictors and an indexed sequence of univariate distributional responses. For each sample $i = 1, \dots, N$, we generate a predictor vector $\boldsymbol{x}_i = (x_{i1}, \dots, x_{ip})^\top \in \mathbb{R}^p$, with $p = 20$. Predictor components are independently sampled as $x_{ik} \sim \text{Uniform}(5,10)$. Only the first six predictors are truly active, forming the active set $S$, while the remaining predictors are nuisance variables. This setup allows us to assess DFR's ability to perform geometry-aware feature selection in high-dimensional settings.

Each sample is observed over $L = 10$ ordered indices (e.g., time points or layers). At each index $l$, the response $Y_{il}$ is a univariate normal distribution whose mean and standard deviation depend only on the active predictors. Specifically, we define predictor-dependent, index-specific mean $\mu_l(\boldsymbol{x}_i)$ and standard deviation $\sigma_l(\boldsymbol{x}_i)$ functions

$$\mu_l(\boldsymbol{x}_i) = \mu_0 + \beta_l \sum_{k \in S} x_{ik}\,;\ \sigma_l(\boldsymbol{x}_i) = \sigma_0 + \gamma_l \sum_{k \in S} x_{ik}$$

with baseline values $\mu_0 = 0.0$ and $\sigma_0 = 0.1$. The coefficients $\beta_l$ and $\gamma_l$ quantify how strongly the active predictors influence the distributional location and spread at index $l$, respectively. Allowing $\beta_l$ and $\gamma_l$ to vary with $l$ induces structured distributional evolution across indices, while keeping the predictor values fixed for each sample. To introduce additional stochastic variability across samples and indices, we perturb these parameters by sampling

$$\mu \sim \mathcal{N}(\mu_l(\boldsymbol{x}_i), v_1), \sigma \sim \text{Gamma}\left(\frac{\sigma_l(\boldsymbol{x}_i)^2}{v_2}, \frac{v_2}{\sigma_l(x_i)}\right)$$

where $v_1 = 0.00025$ and $v_2 = 0.001$. Condition on $(\mu, \sigma)$, the response distribution at predictor $x_i$ and index $l$ is represented through its quantile function evaluated on a common grid $u \in [0.01, 0.99]$:

$$q_l(u \mid \boldsymbol{x}_i) = \mu + \sigma\, \Phi^{-1}(u)$$

with $\Phi^{-1}$denoting the standard normal quantile function. Each sample thus yields a tensor of quantile values $q_i \in \mathbb{R}^{L \times D}$, where $D$ is the number of quantile grid points.

To study different evolution dynamics of distributions across indices, we consider three scenarios that differ only in how the index-specific coefficients $\{\beta_l, \gamma_l\}$ vary with $l$, while keeping the distribution family and stochastic perturbation model fixed.

- **Scenario I (Linear trend)**: Covariate influence increases linearly with the index,

$$\beta_\ell = \beta_{\min} + (\beta_{\max} - \beta_{\min})\frac{\ell - 1}{L - 1}, \gamma_\ell = \gamma_{\min} + (\gamma_{\max} - \gamma_{\min})\frac{\ell - 1}{L - 1}.$$

- **Scenario II (Quadratic trend)**: Covariate influence accelerates at later indices, with

$$g_\ell = \left(\frac{\ell - 1}{L - 1}\right)^2, \beta_\ell = \beta_{\min} + (\beta_{\max} - \beta_{\min})g_\ell, \gamma_\ell = \gamma_{\min} + (\gamma_{\max} - \gamma_{\min})g_\ell.$$

- **Scenario III (Periodic trend)**: Covariate influence varies cyclically across indices,

$$\beta_\ell = \beta_{\text{mid}} + \beta_{\text{amp}}\sin(\theta_\ell), \gamma_\ell = \gamma_{\text{mid}} + \gamma_{\text{amp}}\sin(\theta_\ell),$$

where $\theta_\ell = \frac{2\pi}{L/2}(\ell - 1)$, $\beta_{\text{mid}} = (\beta_{\max} + \beta_{\min})/2$, $\beta_{\text{amp}} = (\beta_{\max} - \beta_{\min})/2$, and analogously for $\gamma_{\text{mid}}$ and $\gamma_{\text{amp}}$. Throughout, we set $\beta_{\min} = 1.0$, $\beta_{\max} = 8.5$, $\gamma_{\min} = 0.3$, and $\gamma_{\max} = 0.8$.

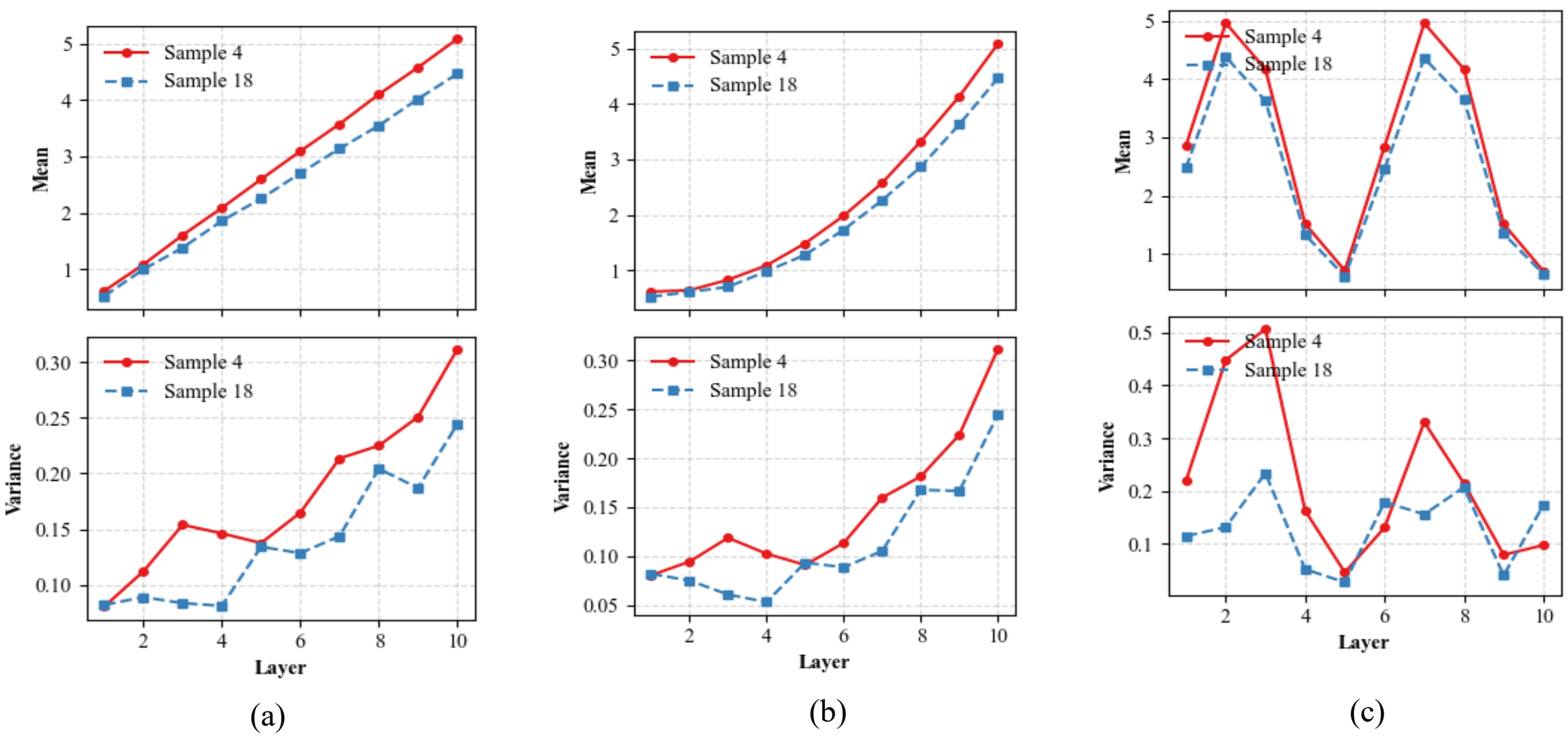

Figure 1. Evolution of the distributional mean and variance under the three simulation scenarios, shown for two samples: (a) Scenario I (Linear); (b) Scenario II (Quadratic); (c) Scenario III (Periodic)

Figure 1 illustrates the index-wise evolution of the mean (top row) and variance (bottom row) of normal distributions under the three simulation scenarios. As a representative, we display results for two representative samples (Sample 4 and Sample 18). In Scenario I (Linear trend), shown in panel (a), both the mean and variance increase approximately linearly with the index. This reflects the linear growth of the index-specific coefficients $\beta_l$ and $\gamma_l$ producing steadily strengthening covariate effects across layers. Differences between samples are preserved across indices, as their relative separation is driven by the same active predictors. Panel (b) corresponds to Scenario II (Quadratic trend), where the covariate influence accelerates at later indices. As a result, both the mean and variance exhibit slow changes at early layers followed by more rapid growth toward higher indices. This setting induces nonuniform smoothness across layers and is designed to assess DFR's ability to adapt to uneven, nonlinear distributional dynamics. Panel (c) shows Scenario III (Periodic trend), in which the covariate effects vary cyclically with the index. Consequently, both the mean and variance oscillate across layers, producing repeating peaks and troughs. This scenario represents a challenging nonmonotone evolution pattern, testing whether DFR can capture structured but non-smooth dynamics while still borrowing strength across neighboring indices. Overall, Figure 1 visually confirms that the three scenarios generate distinct, controlled forms of distributional evolution in both location and spread, providing a flexible testbed for evaluating DFR's ability.

### *4.2 Benchmark Models*

Our primary method is DFR-FS-RS (Dynamic Fréchet Regression with Feature Selection and Roughness Smoothing). This approach first estimates an active predictor set $\hat{S}$ via sparse metric learning and then refits the DFR model using the selected predictors $x_{\hat{S}}$, with the roughness parameter $\lambda$ tuned to control the degree of cross-index smoothing. To disentangle the effects of

(i) feature selection and (ii) layer-wise roughness smoothing, we consider the following DFR variants:

- **DFR-Base** (no feature selection, no smoothing): uses all predictors with $\lambda = 0$;
- **DFR-FS** (feature selection only): uses $x_{\hat{S}}$ with $\lambda = 0$;
- **DFR-FS-RS** (feature selection and smoothing; full model): uses $x_{\hat{S}}$and tunes $\lambda$.

We benchmark these methods against two established alternatives: Global Fréchet Regression (GFR) and Function-on-Scalar Regression (FoSR). In FoSR, probability distributions are represented via their quantile functions and treated as Euclidean functional data, thereby ignoring the intrinsic geometric structure and constraints of the probability space. All methods are compared under the same simulated data and identical train–test splits.

### *4.3 Evaluation Metrics*

For each simulated dataset, the 50 samples are randomly split into training (90%) and testing (10%) sets. Each model is evaluated under each scenario using 25 Monte Carlo replications. In each replication, we record: (i) feature selection performance of the proposed DFR-FS-RS method, including precision, recall, F1-score, and accuracy over 20 predictors; (ii) prediction accuracy on the test set of benchmark methods and DFR variants. Prediction accuracy is quantified using the mean squared prediction error (MSE) defined as $\mathrm{MSE} = \frac{1}{N_{test}L} \sum_{i \in test} \sum_{l=1}^{L} d^2\left(Y_{il}^{true}, Y_{il}^{pred}\right)$. 2-Wasserstein distance is used for distributional regression methods and the Euclidean distance is used for FoSR. To quantify uncertainty across Monte Carlo replications, we report the standard deviation of the MSE (and analogously for feature selection metrics) across replications.

### *4.4 Results on Feature Selection*

This section summarizes the feature selection performance of the proposed DFR-FS-RS method under the three simulation scenarios. Table 2 reports the mean precision, recall, F1-score, and accuracy of feature selection across 25 Monte Carlo replications, along with their standard deviations. The results show that the proposed method consistently identifies all truly active predictors in every scenario, achieving perfect recall (mean = 1.00, standard deviation = 0.00). Precision ranges from 0.65 to 0.77, indicating that occasional false positives occur—i.e., a small number of nuisance predictors are selected in some replications—but the selected predictor sets remain highly informative. Overall, the resulting F1-scores are strong, with values of 0.83 for the Linear scenario, 0.77 for the Quadratic scenario, and 0.86 for the Periodic scenario. The Periodic scenario achieves the best balance between retaining all active predictors and suppressing irrelevant ones, a pattern that is also reflected in the accuracy metric, which is highest under this scenario. These findings suggest that the geometry-aware metric-learning stage of DFR-FS-RS adopts a conservative selection strategy: it prioritizes retaining all active predictors (perfect recall) while permitting mild over-selection when the underlying signal structure is more complex, as in the Quadratic scenario. Importantly, the presence of a small number of nuisance predictors is tolerable in the proposed method, since the subsequent refitting and layer-wise smoothing stages further stabilize estimation and prediction.

Table 2. Mean and standard deviation of feature selection accuracy by DFR-FS under the Linear, Quadratic, and Periodic scenarios over 25 Monte Carlo replications.

| Scenario | Precision | Recall | F1 Score | Accuracy |
|---|---|---|---|---|
| Linear | 0.73 (0.20) | 1.00 (0.00) | 0.83 (0.14) | 0.85 (0.14) |
| Quadratic | 0.65 (0.20) | 1.00 (0.00) | 0.77 (0.14) | 0.80 (0.16) |
| Periodic | 0.77 (0.19) | 1.00 (0.00) | 0.86 (0.13) | 0.88 (0.13) |

### *4.5 Results on Prediction Accuracy*

Table 3 reports the mean test MSE and corresponding standard deviation for each method across the three simulation scenarios, averaged over 25 Monte Carlo replications. Several clear patterns emerge. First, the distributional regression approaches—Global Fréchet Regression (GFR) and the proposed Dynamic Fréchet Regression (DFR) variants—substantially outperform Function-on-Scalar Regression (FoSR) in all scenarios. This result highlights the importance of explicitly accounting for the geometric structure and constraints of probability distributions when modeling distribution-valued responses; treating distributions as Euclidean functional data leads to large errors and instability.

Second, all Dynamic Fréchet Regression variants (DFR-Base, DFR-FS, and DFR-FS-RS) consistently achieve lower test MSE than Global Fréchet Regression. This improvement demonstrates the benefit of incorporating index-aware weighting, which enables DFR to exploit cross-index dependence and produce more accurate index-specific predictions than the global, index-agnostic weighting used by GFR. The advantage of DFR is particularly pronounced under the Periodic scenario, indicating that the proposed method is robust to complex, non-monotone distributional dynamics across indices.

Third, incorporating feature selection further improves predictive performance. Both DFR-FS and DFR-FS-RS outperform DFR-Base, with the largest gains observed under the Quadratic scenario, where covariate effects evolve more rapidly across indices. This finding suggests that removing nuisance predictors enhances prediction accuracy, especially when the underlying signal structure is nonlinear and heterogeneous. Finally, the performance of DFR-FS and DFR-FS-RS is nearly identical across all scenarios, indicating that while feature selection plays a critical role, additional roughness smoothing provides only marginal gains in these settings.

Table 3. Mean MSE and standard deviation (in parentheses) of prediction accuracy of different methods under the Linear, Quadratic, and Periodic scenarios, averaged over 25 Monte Carlo replications.

| Model | Scenario I (Linear) | Scenario II (Quadratic) | Scenario III (Periodic) |
|---|---|---|---|
| **DFR-FS-RS** | **0.030 (0.016)** | **0.626 (0.711)** | **0.031 (0.017)** |
| DFR-FS | 0.031 (0.017) | 0.626 (0.711) | 0.031 (0.017) |
| DFR-Base | 0.034 (0.020) | 0.668 (0.786) | 0.034 (0.019) |
| FoSR | 5.112 (7.680) | 1.523 (2.597) | 8.145 (11.715) |
| GFR | 1.086 (0.856) | 4.357 (3.310) | 7.453 (4.824) |

## 5. Case Study in Additive Manufacturing

We apply the proposed Dynamic Fréchet Regression (DFR) framework to model and predict dynamically evolving distribution-valued responses arising in an additive manufacturing process. This application naturally fits the DFR setting: static process parameters act as scalar predictors, while the responses are sequences of probability distributions that characterize both within-layer variability and between-layer evolution of melt-pool behavior. Modeling these distributional relationships is critical, as process parameters influence not only the typical melt-pool response but also its variability and uncertainty, which are closely linked to defect formation and part quality (Karthikeyan et al. 2025; Tang et al. 2017; Yang et al. 2016). By explicitly learning how process parameters govern layer-wise melt-pool distributions, DFR provides actionable insight for process monitoring, control, and optimization, enabling informed adjustment of process settings to achieve stable thermal behavior and improved build consistency across layers.

### *5.1 Data collection and preprocessing*

The dataset was collected from a real-world Directed Energy Deposition (DED) process, illustrated in Figure 2(a) (Karthikeyan et al. 2025). It consists of 25 builds produced under an augmented central composite design (CCD) involving three process parameters: laser power (LP), scanning speed (SS), and inter-layer dwell time (DT). The CCD includes factorial, axial, center, and extreme-factorial points, providing broad coverage of the process space and enabling estimation of main effects, two-factor interactions, and curvature (quadratic) effects. All printed samples are shown in Figure 2(b). Each build consists of 16 sequential layers.

During fabrication, melt-pool behavior is monitored using a high-speed thermal imaging system (Figure 2 (a)), which records a time-ordered sequence of thermal images. Within each layer, the images capture melt-pool behavior at different spatial locations along the toolpath, while images across layers reflect the evolution of thermal dynamics with build height. Each thermal image is a $200 \times 200$matrix of temperature values, associated with a spatial location and a layer index. Images are aligned by location and layer to enable consistent analysis. From the thermal image corresponding to sample $i$, layer $l$, and spatial location $s$, we extract two physically interpretable melt-pool features (Akbari et al. 2022; Wang et al. 2023; Zhu et al. 2023):

(1) **Peak melt-pool temperature**, $T_{ils}^{\max}$, defined as the maximum pixel temperature in the image; and

(2) **Melt-pool area**, $A_{ils}$, defined as the area of the region whose temperature exceeds a threshold of 1650 °C, delineating the melt-pool boundary.

An example thermal image and the extracted features are shown in Figure 2(c). To characterize within-layer variability, we model these features probabilistically. For each sample $i$ and layer $l$, the collections $\{T_{ils}^{\max}\}_s$ and $\{A_{ils}\}_s$ are aggregated into empirical distributions, denoted by $T_{il\cdot}^{\max} \sim Y_{il}^{(T)}, A_{il\cdot} \sim Y_{il}^{(A)}$, where $Y_{il}^{(T)}, Y_{il}^{(A)} \in \Omega$, the space of probability distributions. Thus, for each build, both melt-pool temperature and area are represented as layer-indexed distributional trajectories, capturing variability within layers and systematic evolution across layers.

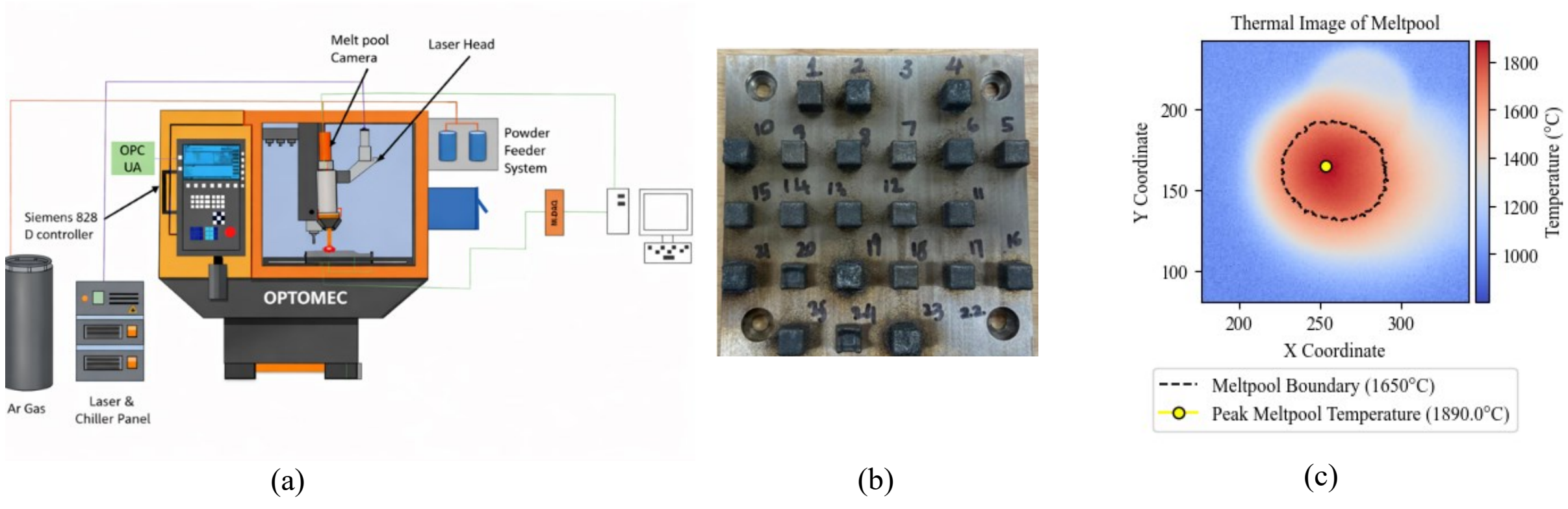

(a) (b) (c)

Figure 2. (a) DED system, (b) 25 printed samples c) Thermal Image of the meltpool for sample.

### *5.2 Modeling setup and tuning*

To model the effects of process parameters on melt-pool behavior within and across layers, we apply the proposed DFR-FS-RS method and compare its performance with benchmark models, including Global Fréchet Regression (GFR) and Function-on-Scalar Regression (FoSR). The predictor set includes the three process parameters and their second-order transformations to capture nonlinear effects and interactions. Specifically, the candidate predictors consist of the main effects $(\mathrm{LP}, \mathrm{SS}, \mathrm{DT})$, quadratic terms $(\mathrm{LP}^2, \mathrm{SS}^2, \mathrm{DT}^2)$, and pairwise interactions $(\mathrm{LP} \times \mathrm{SS}, \mathrm{LP} \times \mathrm{DT}, \mathrm{SS} \times \mathrm{DT})$. Two separate models are fitted for the two response types: the layer-wise distributions of melt-pool area and peak melt-pool temperature. To obtain a parsimonious and interpretable model, we adopt the proposed two-stage DFR-FS-RS workflow, which combines geometry-aware feature selection with index-wise smoothing. In the first stage, sparse metric learning is used to identify the active subset of predictors by learning a sparse similarity metric that governs the covariate-dependent weighting mechanism. The sparsity parameter $\mu$ is selected using leave-one-out cross-validation (LOOCV) to balance predictive accuracy and model complexity. In the second stage, the DFR model is refit using only the selected predictors, and the roughness parameter $\lambda$, which controls the degree of information sharing across layers, is tuned using LOOCV. Given the dataset of $N = 25$ builds, we perform $N$ cross-validation iterations. In each iteration, one build is held out as the test sample, and the model is trained on the remaining $N - 1$ builds. Using the trained model, we predict the full sequence of layer-wise distributions for the held-out build. The prediction error for the held-out build is quantified as the average squared 2-Wasserstein distance between the predicted and observed distributions across all layers,

$$MSE_i = \frac{1}{L}\sum_{l=1}^{L} d^2\left(Y_{il}, \hat{Y}_{il}\right),$$

where $Y_{il}$ denotes the observed distribution for build iii at layer $l$, $\hat{Y}_{il}$ denotes the corresponding prediction obtained from the model trained without build $i$, and $d^2(\cdot,\cdot)$ is the 2-Wasserstein distance. This procedure produces one prediction error value for each held-out build. The average mean squared error and the corresponding standard deviation are computed across all held-out builds. The optimal tuning parameters $\mu$ and λ are selected as those that minimize the average LOOCV MSE. For benchmarking, we also fit Global Fréchet Regression (GFR) and Function-on-Scalar Regression (FoSR). The same LOOCV procedure is applied to all competing methods to ensure a fair and consistent comparison of predictive accuracy and model stability.

### *5.3 Results*

Table 4 summarizes predictive performance for both response types—peak melt-pool temperature and melt-pool area distributions—using test MSE based on the squared 2-Wasserstein distance. Across both responses, DFR-FS-RS consistently achieves the lowest average test MSE and substantially reduced variability, indicating more accurate and more stable distributional predictions than Global Fréchet Regression (GFR) and Function-on-Scalar Regression (FoSR). The improvement is particularly pronounced for melt-pool area, where DFR-FS-RS reduces both the mean error and error dispersion relative to the benchmarks. This aligns with DFR's ability to borrow strength across layers while preserving the intrinsic geometry of distributional responses. Feature selection results of DFR-FS-RS are summarized in Table 5. For both melt-pool area and peak temperature, the proposed method consistently identifies the quadratic term of laser power as a key predictor, in agreement with prior findings that laser power is the dominant driver of melt-pool thermal behaviour (Zhang et al. 2021). The selection of quadratic laser-power effects indicates a nonlinear relationship between energy input and melt-pool response, reflecting saturation and thermal accumulation phenomena. For melt-pool area, interaction terms involving laser power—specifically LP×SS and LP×DT—are also selected, highlighting how the influence of laser power is modulated by scanning speed and dwell time in shaping geometric variability. Overall, this case study demonstrates that DFR provides a flexible, interpretable, and geometry-preserving framework for modeling dynamic distribution-valued responses in real manufacturing systems.

Table 4. Average MSE and its standard deviation of predicting meltpool feature distributions via different models in leave-one-out cross validation.

| Model | Peak Meltpool Temperature | Meltpool Area |
|---|---|---|
| **DFR-FS-RS** | **0.037 (0.016)** | **0.034 (0.019)** |
| **FoSR** | 0.044 (0.032) | 0.042 (0.052) |
| **GFR** | 0.044 (0.027) | 0.047 (0.045) |

Table 5. Feature selection results of DFR-FS-RS for different meltpool features.

| Response | Selected Features |
|---|---|
| **Peak Meltpool Temperature** | $\mathrm{LP}^2$ |
| **Meltpool Area** | $\mathrm{LP}^2$, LP×SS, LP×DT |

## 6. Conclusion

This paper proposed Dynamic Fréchet Regression (DFR) for modeling and predicting sequences of autocorrelated probability distributions from scalar process parameters, motivated by the stochastic and evolving nature of melt-pool behavior in Directed Energy Deposition (DED). In contrast to conventional regression and functional regression methods that rely on Euclidean structure, DFR performs prediction through weighted Fréchet means in a metric space of distributions, thereby preserving the geometry induced by distances such as the 2-Wasserstein metric. By incorporating index-aware weighting derived from a roughness-penalized function-on-scalar analogue, DFR produces index-specific distributional predictions while borrowing information across neighboring layers, leading to improved stability and interpretability.

The paper also introduced a geometry-aware feature selection procedure for dynamic distributional regression. By learning a sparse predictor metric that governs similarity weights, the proposed approach enables variable selection without requiring explicit regression coefficients for distribution-valued responses. Simulation studies demonstrate that the resulting two-stage procedure—feature selection followed by refitting with tuned roughness smoothing (DFR-FS-RS)—consistently improves predictive accuracy over Global Fréchet Regression (GFR) and Function-on-Scalar Regression (FoSR), while effectively identifying active predictors in the presence of nuisance variables.

The methodology was further illustrated using a DED case study with 25 builds and 16 layers. For both peak melt-pool temperature and melt-pool area, DFR-FS-RS achieved the lowest test error and reduced variability relative to the benchmark methods. The selected models were interpretable and consistent with physical understanding of the process: quadratic laser-power effects dominated peak-temperature distributions, while melt-pool area additionally depended on interactions between laser power, scanning speed, and dwell time. These results highlight the practical value of modeling distributional responses, rather than summary statistics, for understanding process behavior and variability.

Overall, DFR provides a practical and interpretable framework for regression with dynamic, distribution-valued responses. Although motivated by additive manufacturing, the proposed methodology is applicable to a wide range of problems involving indexed distributional data, including process monitoring, reliability analysis, and other industrial and engineering applications where quantifying and predicting variability is essential.

**Acknowledgement**: This work was funded by award 70NANB23H028 from the National Institute of Standards and Technology (NIST) and supported by Advanced Manufacturing Institute (AMI) at the University of Houston.

**Data availability statement**: The data that support the findings of this study are available from the corresponding author, YL, upon reasonable request.